# Preparation of large-scale ring carbon nanotube networks and a general growth mechanism for carbon nanotubes

Zhi-An Ren[*] and Jun Akimitsu

*Department of Physics and Mathematics, Aoyama-Gakuin University,*

*Sagamihara, Kanagawa, 229-8558, Japan*

**Abstract**:

Large-scale fully interconnected ring carbon nanotube (CNT) networks were first prepared using thermo chemical vapor deposition in nano-channel network templates of porous anodic alumina. This conductive CNT network film consists of billions of nanotube segments with a single layer, and could be made as large as the centimeter size of the template, with a uniform two-dimensional ring topological structure. These CNT networks could be grown from tiny Fe/Co catalyst particles which were electro-deposited into the templates, or totally without additional metallic catalysts. The common interconnected ring topology suggests a both-tip growth mechanism in which the growth of every CNT should occurs at both ends by the incorporation of carbon clusters until they connect into other CNTs. Comparing the different morphology of CNT networks grown with and without catalyst particles, we found that the local concentration of carbon clusters should play the key role for the continued growth of CNTs.

**Keywords**: Carbon nanotubes; Chemical vapor deposition; Pyrolysis; Atomic force microscopy; Electron microscopy.

---

[*] Corresponding author. Fax: +81-42-759-6287. Email address: renzhian@phys.aoyama.ac.jp (Zhi-An Ren)

## 1. Introduction

One of the most intriguing problems in nanoscience is to understand the microscopic growth mechanism of these wonderful tubular structural carbon allotropes [1], and then realize the controllable growth of carbon nanotube (CNT) patterns for utilizations. Till now experimental techniques have been widely developed and CNTs could be produced with various methods and very flexible environments (from higher than 3000$^{o}$C of arc discharge, laser ablation to as low as 500$^{o}$C of chemical vapor deposition (CVD) methods) [2-7]. Because of their unique quasi-one-dimensional structures, CNTs have very different chirality, diameters, layers and physical properties, and the growth conditions and behaviors also differ widely. Different growth mechanisms have been proposed to explain the underlying initiating process and growing dynamics [7-10], and most of them are focusing on the metallic catalyst-assisted growth, in which the precipitation or diffusion of carbon atoms (produced from the reaction of hydrocarbon feedstock with catalyst particles) at the catalyst surface is believed to provide the continuing growth of CNTs, while all those proposed mechanisms are still highly controversial because of lacking of experimental proofs and incapable of explaining for all growing behaviors. Hereinafter we report a new morphology of CNTs, the fully interconnected two-dimensional ring network CNTs, which were grown by low temperature CVD from the well-prepared nano-channel network template in porous anodic alumina [11-13]. Based on the special macroscopic ring topological structure, we approached the microscopic formation processes for these CNT networks and a general growth mechanism for multiwall CNTs (MWNTs), in which all these CNTs should strictly grow in a both-tip mechanism, with ends open and growing forwards in both directions by the incorporation of carbon clusters from both ends. The discovery of this network structure of CNTs may greatly enhance the development of applications for nanotechnology, while the understanding for the underlying microscopic growth mechanism is crucial to the controllable growth of CNTs and designable two-dimensional and three-dimensional nanotube architectures.

## 2. Experimental methods

The synthesis of the CNT network started from the preparation of interconnected nano-channel network templates in porous alumina, which is depicted in Fig. 1. Briefly, at first a layer of porous alumina film with a uniform separated pore structure (Fig. 1a) was prepared on an aluminum sheet by anodization in acid solution under a stable environment [11]. After that the anodizing condition was changed to form a very thin layer of porous film with much narrower and denser pores beneath the previous oxide film (Fig. 1b). Because the oxide walls between these narrower pores are much thinner in this new layer, a subsequent appropriate acid etching process could etch through them and connect these pores together by transverse channels to produce an interconnected planar nano-channel network template at the bottom of porous alumina layer (Fig. 1c). By electro-depositing tiny metallic catalyst particles into part of these pores' bottom (or not) [13], CNTs could be catalytically (or non-catalytically [14]) grown along these nano-channels to form a network from a simple CVD process by pyrolysis of acetylene ($C_2H_2$) which flowed into the channel-template through vertical pores. After removed the surface oxide layer, CNT networks could be observed by electron microscopy. Fig. 1d is a transmission electron microscopy (TEM) cross-section image of the alumina template, which clearly displays that the pores are connected together by etched channels at the bottom, as indicated by arrows. And the sample was immersed in saturated $HgCl_2$ solution to remove the aluminum base to expose the structure of the alumina template' backside. An Field-Emission Scanning Electron Microscopy (FESEM) image shows the typical convex structure of the backside of barrier oxide layer (Fig. 1e), which indicates that the alumina barrier was not etched through and CNT networks were grown totally in alumina template without contacting with the base aluminum. In Fig. 1f, the schematic inner structure of the nano-channel network template is depicted, the original pores and channels connecting them are marked with 'A' and 'B'. For templates used in this paper, the diameters of pores and channels are about 100 nm and 20 nm respectively determined by FESEM and TEM measurements.

Typically, the nano-channel network template was fabricated through three steps. At first, an electrochemical-polished high purity aluminum sheet (0.5 mm in thickness) was anodized in 0.3 mol/L oxalic acid solution under $12^oC$ by 40 Volts for 20 minutes (which would make a 2 μm thick porous alumina film on the surface). Then the anodizing voltage was slowly decreased to 15 Volts in 3 minutes, and kept for another 2 minutes to grow smaller pores. Finally the template was etched by flowing phosphoric acid (5%) for 12 minutes under 40$^{o}$C to make an interconnected nano-channel network beneath the porous film. For catalytic growth, tiny Fe/Co bimetal catalyst particles were deposited into the bottom of some pores by pulsed electro-deposition method (the particle size should be controlled to be smaller than the diameter of nano-channels so that CNTs could grow horizontally.) [13]; for non-catalytic growth, no metallic catalysts were used (CNTs grown in alumina pores without catalysts have been reported before [14]). The final templates were put into a sealed quartz tube furnace on a small quartz boat and heated to 620$^{o}$C under vacuum (less than 1 Pa). After 5 minutes reduction by hydrogen gas, the growth of CNT networks was performed by pyrolysis of acetylene for 20 minutes with a flow of mixed gas ($C_2H_2$: Ar ~ 1: 9) at 150 standard cubic centimeter per minute, the pressure inside the quartz tube was maintained at 200 Pa by a high speed rotary pump during growing. After cooled down in vacuum, the surface alumina layer on the template was carefully removed by phosphochromic acid [11] and the CNT network adhered on aluminum surface would be exposed. Samples were characterized by FESEM (JEOL JSM-6301F), atomic force microscopy (AFM, SHIMADZU SPM-9500) and TEM (JEOL JEM-4010). This type of nano-channel network template has been successfully produced by anodization in sulfuric acid and oxalic acid aqueous solution with a starting voltage from 15 Volts to 60 Volts. And CNT networks with different density and diameters could be prepared by these templates. Actually, the morphology of CNT networks is difficult to control because of the sensitive etching process and catalyst-deposition which is strongly affected by the nonuniform barrier layer.

Because high density pores acted as gas pipelines, the CNT network could grow very quickly in spite of the template size, and could be formed completely even in 2 minutes reaction for catalytic growth in our experiments.

## 3. Sample characterizations

In Fig. 2, typical CNT networks are exhibited by FESEM images for several different samples. For networks grown from catalysts, all CNTs are interconnected with both tips to form a whole ring network. The network-density and diameters of these nanotubes are variable for different samples, but in one sample, uniform network structure could be formed in the whole template with as large as 5 square centimeters size (the best sample we obtained, as in Fig. 2c). While the uniform CNT networks could only be produced in the rightly-prepared templates, most of samples have nonuniform structures as in Fig. 2d, which should be caused by the accumulation of catalysts along the excess etching tracks (where the much thinner barrier layer caused the bigger accumulated catalyst particles). The CNT segments in a network have different lengths varying under several micrometers even in a small area, which is due to the random depositing positions of catalyst particles. Most of junctions in the network have a 3-branch structure, while some of them can connect to 4 or 5 CNT segments. The junction density varies from 1-10% of the density of pores, corresponding to the sparse density of catalyst particles, and some catalyst particles can be observed in junctions, as indicated by arrows in Fig. 2a. A CNT network was scratched by a needle to observe the broken behavior (Fig. 2e), which shows the network is stretched by force and very flexible, and after scratched the CNTs at the broken edge shrinked together. For CNT networks grown without metallic catalysts (Fig. 2f), usually they have much higher junction density (near 50% of pores) and shorter nanotube segments. These nanotubes are not fully interconnected, lots of tips are floating, and some separated very short CNTs also exist; but some ring networks can still be observed as marked by arrows. The absence of catalysts should need different growth mechanism for these networks, while the common ring topological structure will bring us to find the general growth mechanism for all these CNTs, as what we will discuss later. From the template structure and the CNT network morphology, it can be concluded that for catalytic growth, catalyst particles should be randomly deposited on the bottom of templates with a very sparse density, and CNTs only grew from catalysts to form a network. Without catalysts, CNTs should initiate nucleation from the inside tubular structure of nano-channels, which made them have similar diameters and densities. Actually the non-catalytic growth of CNT networks was very difficult (most of samples could not grow CNTs at all, and the reason is still not clear), which may explain that why no non-catalytically grown CNTs accompanied with catalytic growth.

The CNT networks were also characterized by AFM, in which the topography of the sample surface was detected, and from which the diameters of these CNTs could be obtained more precisely. Fig. 3a shows a typical AFM image, which displays the same catalytically grown CNT network structure with SEM images. The diameters of these nanotube segments were calculated from the height profile of crossing lines, as shown by Fig. 3b. By the data obtained from several hundreds of nanotube segments on different samples, we found that the diameters of CNTs may vary from 10 nm to 30 nm for different samples (which are all typical MWNTs), but for one sample with centimeter size and billions of nanotube segments, almost all nanotubes have similar diameters with a narrow distribution. As in fig. 3a, almost all these nanotube segments in the network have diameters around 15 nm. Since the catalyst particles produced by electro-deposition have uncontrollable and nonuniform sizes, the similar diameters of all these CNTs may indicate that the diameters of CNTs might not directly depend on the size of catalyst particles.

To separate the CNT network from template, small holes were drilled through the as-prepared sample around edges, and gold filaments (20 μm in diameter) were put through these holes and fixed at a plastic frame. The sample was immersed in diluted NaOH solution by overnight. The aluminum base and alumina surface layer could be completely dissolved and the CNT network would be suspended in air by gold filaments. This CNT network film is totally transparent because of its thin thickness and low density (typically, 1 gram of this CNT network could expand to larger than 1000 square meters according to the density calculation). The network film is conductive and the room temperature resistance is usually several mega ohms from two gold electrodes between several millimeters.

## 4. Discussions and conclusions

The catalytic growth of CNTs is currently believed to initiate from catalyst particles (similar as seed-growth), and continue growing at the catalyst tip by the incorporation of carbon atoms from precipitation or diffusion at the catalyst surface either in root-growth mode or tip-growth mode, with the other tip open or closed by fullerene-like cap while growing [7]; although there is still no explicit evidences about how does CNT grow at the catalyst tip and what happens at the other tip during the high temperature reaction, and CNTs can even grow without catalyst under some circumstances. Usually the growing mechanism was characterized by several picked nanotubes from quenched or *in-situ* growth [15] with high resolution microscopy. But because of various preparing methods and growing behaviors exist, those proposed mechanisms are incapable to be verified and applied for all CNTs. Here we start to analyze the formation process and growth mechanism for catalytically grown CNT networks based on above experimental results and the ring topology. The schematic interconnected ring network is illustrated in Fig. 4a, which is similar with catalytically grown CNT networks without floating tips. As we know, CNTs grow from points to segments, and because the network only consists of a single layer of CNTs, when moving nanotube tips meet other nanotubes (either wall or tip), they should stop growing and connect together but not cross each other. However, if we presume that every CNT grows from a fixed point (catalyst particle) to one direction only, just like drawing line segments from many fixed points to random directions and stopping when meeting each other, we will find that only tree networks could be formed, as illustrated in Fig. 4b, and no interconnected ring topology could be produced.

Here we start to consider that both tips of every CNT can move forwards when growing, and eventually they will meet their partners and connect together. From this scenario the ring network will finally be formed since all tips will connect to other CNTs and no floating tips can be left (as Fig. 4a). Because most of the moving tips will meet other CNTs' walls rather than tips (the much larger area of walls give them more probability, this also explains the majority of 3-branched junctions), and the walls have stable local surface structures, to connect together, the ends should be open when growing so that the active dangling bonds of carbon atoms at the end edge are able to connect onto the wall surface where they meet. And on the other hand, only when the ends are open, they can be inserted with carbon clusters to continue growing. This scenario is depicted in Fig. 4c. A nanotube grows with a catalyst particle at one end and with another end open, carbon clusters could be inserted from both ends so that the nanotube will grow along both directions. This picture also suggests that without catalysts CNTs still could grow once they are initiated with open ends (as non-catalytic CVD growth, and this should also be similar with the widely used non-catalytic arc-discharge growth of MWNTs). In fact, this mechanism could be verified more carefully from Fig. 1 by the statistical probability distribution of catalyst particles and the number of junction-branches in the network. Because catalyst particle stays at only one tip of the CNT segment, and most of the tips connect to other nanotubes' walls, thus only about half of the 3-branch junctions will contain catalysts, and only when two tips meet together, there may be a catalyst embedded in the center of one nanotube. The probability for two or more nanotube tips to meet together is quite low and depending on the network density and template structure, which makes the 4-branch or 5-branch junctions very rarely.

One may consider that some CNTs will follow this both-tip growth mechanism, but other nanotubes still grow only from the catalyst tip either in root-growth mode or tip-growth mode, because this also could form a fully interconnected ring network without floating tips logically. In fact, if we consider that some tips will stop growing by some perturbation caused cap closure or catalyst poisoned effects, and these tips will stay at their pores and wait for other tips to voluntarily find them and connect together. However, because of the low density of tips (as low as to 1% of pores), it will be very difficult for another moving tip to find this immobile tip in so many pores, and to a macro number of stopped tips (millions to billions), that would be impossible for all of them to be connected. This will result in a network with many floating tips. Another concern is that the open tip of a CNT may move forwards by pushing from the catalyst side but not actively growing, while under this consideration this kind of 'push-moving' tip will stop moving when this nanotube is connected and held by another nanotube, which will become similar situation as above and result in many floating tips. Therefore to form a network without floating tips, it is essential that all tips of CNTs should grow and move forwards actively, until they meet another nanotube and connect together. Because usually CNTs can grow to several micrometers and even much longer without stopping, the ring network can be easily produced from this both-tip growth mechanism. We note that from this picture the above template structure is not really necessary, a limited planar space (which can confine CNTs growing in-plane) with sparse dispersed catalyst particles should be enough to produce CNT networks.

Now we discuss about the formation process of non-catalytically grown CNT networks which contain many floating tips (Fig. 1f). As we discussed above, these CNTs should initiate from the tubular structure of transverse nano-channels. After that, the same both-tip growth mechanism (but two open ends without catalyst particles) should be activated because ring networks exist. Somehow, most of them stopped growing before they met other nanotubes, which resulted in a network structure as in Fig. 1f. The interrupting of the growth should be ascribed to the absence of catalysts because that's the only difference with catalytically grown networks. Now we rethink that in catalytically grown networks the open nanotube tips without catalysts could also grow for enough long distances, which will need lots of active carbon clusters floating nearby. As we know the transition metallic particles can greatly enhance the dissociation of hydrocarbon molecules into carbon clusters floating outside of catalyst particles under high temperature, which can be seen by the accumulated amorphous carbon on the inner surface of quartz tube. Therefore for catalytically growing networks, the concentration of carbon clusters should be quite high so as to enable all nanotubes continuing growing for enough long distances; but for non-catalytically growing samples, the insufficient concentration of carbon clusters would cause nanotubes to stop growing very quickly, which resulted in many floating tips.

Then we can summarize the growth process of these CNT networks. Under high temperature or with the assistance of catalysts, hydrocarbon molecules will dissociate into carbon clusters, and these carbon clusters will connect together to make the initial forms of tubular structures around catalyst particles or in some nano-channels (while the molecular dynamics of these initiating processes is still unclear). These CNTs are initiated with open ends, which makes them can grow towards both directions actively by incorporation of carbon clusters from both ends, and connect together to make a whole network when they meet each other.

Theoretically, there are other growth modes or possibilities for CNTs that could also form this interconnected ring network, such like that one nanotube will split into several branches when growing [16-17], or two or more nanotubes could grow out from one fixed catalyst particle [18], or new nanotubes could grow on the walls of other nanotubes [19-20], etc. But based on the morphology of the CNT networks and inner template structure, those mechanisms are neither practical here nor could they form an interconnected network free of floating tips. Because the fully interconnected ring network proved that all CNTs should abide by this both-tip growth mode, and these experiments performed at a relatively low temperature (higher temperature is believed to be more favorable for keeping ends open), this should be a more general growth mechanism for all MWNTs with similar diameters. Besides of the vast application foreground, this ring network structure also provides a universal platform for the study of growth mechanisms for smaller diameter multiwall or singlewall CNTs and other materials.

**Acknowledgements:**

The authors acknowledge Dr. T. Muranaka, A. Miyamura, Prof. J. Haruyama, and Mr. N. Kobayashi, I. Shibayama, S. Nakamura, K. Hasegawa, H. Yamaguchi for experimental assistances, and Prof. H. Shinohara, S. H. Lee, and Dr. J. I. Sohn, A. Y. Park for helpful discussions. This work was supported by the 21$^{st}$ century COE program, "High-Tech Research Center" Project for Private Universities: matching fund subsidy from MEXT, and a Grand-in-Aid for Scientific Research on Priority Area from MEXT.

**Figure captions:**

Fig. 1: The structure of porous anodic alumina template. (a) Schematic hexagonal pore structure of porous alumina, the pore cell parameter $D_C$ and pore diameter $D_P$ have approximate relationships with anodizing voltage $V$ ($D_C$ (nm) = $n$ (nm/Volt) * $V$ (Volt) = $a$ * $D_P$ (nm)), and the barrier layer thickness $L_B$ is about half of $D_C$; constant $n$ and $a$ vary with different acid solution and anodizing temperature, and pore length $L_P$ is proportional to anodizing time. These give us a convenient way to increase or decrease pore diameters and densities by changing $V$ when keeping anodizing to get a desired structure; (b) Schematic cross-section view of template structure after decreasing $V$ to a lower value $V_2$ for a short time; (c) Schematic drawing for the etched template, with nano-channels connecting pores together; (d) TEM observation for the cross-section of the final template (scale bar: 200 nm), where the arrows indicate the etched channels that connect through pores, same with (c); (e) FESEM image for the backside of the final template with convex structure after removing aluminum base (scale bar: 1 μm); (f) Schematic template structure of pores (point A) interconnected by etched nano-channels (point B).

Fig. 2: Typical FESEM images for CNT networks with various morphology. Fe/Co catalysts are used for the growth of sample (a), (b), (c), (d) and (e), while (f) is grown without catalyst. Arrows in (a) indicate some clear embedded catalyst particles, and arrows in (f) indicate some connected ring networks. The broken behavior of the CNT network scratched by a needle is shown in (e). (Scale bars, a: 500 nm; b: 500 nm; c: 5

μm; d: 5 μm; e: 2 μm and f: 1 μm).

Fig. 3: AFM characterization of CNT networks. (a) AFM image of a CNT network (scale bar: 2 μm); (b) the diameter measurements from the height profile of lines crossing nanotubes (from top to base of peaks, around 15 nm).

Fig. 4: Schematic representations for the formation of CNT networks. (a) Schematic interconnected ring network from both-direction growth; (b) Schematic tree networks from one-direction growth; (c) Schematic drawing for the both-tip growth mode for a nanotube grown from a Fe/Co catalyst particle by incorporation with carbon clusters from both ends.

Fig. 1

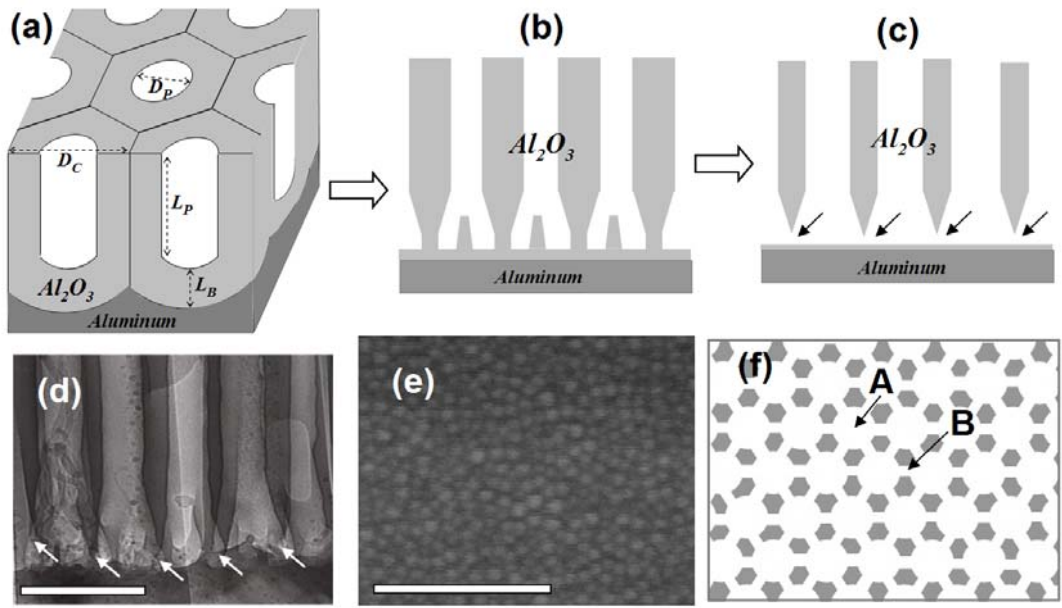

Fig. 2

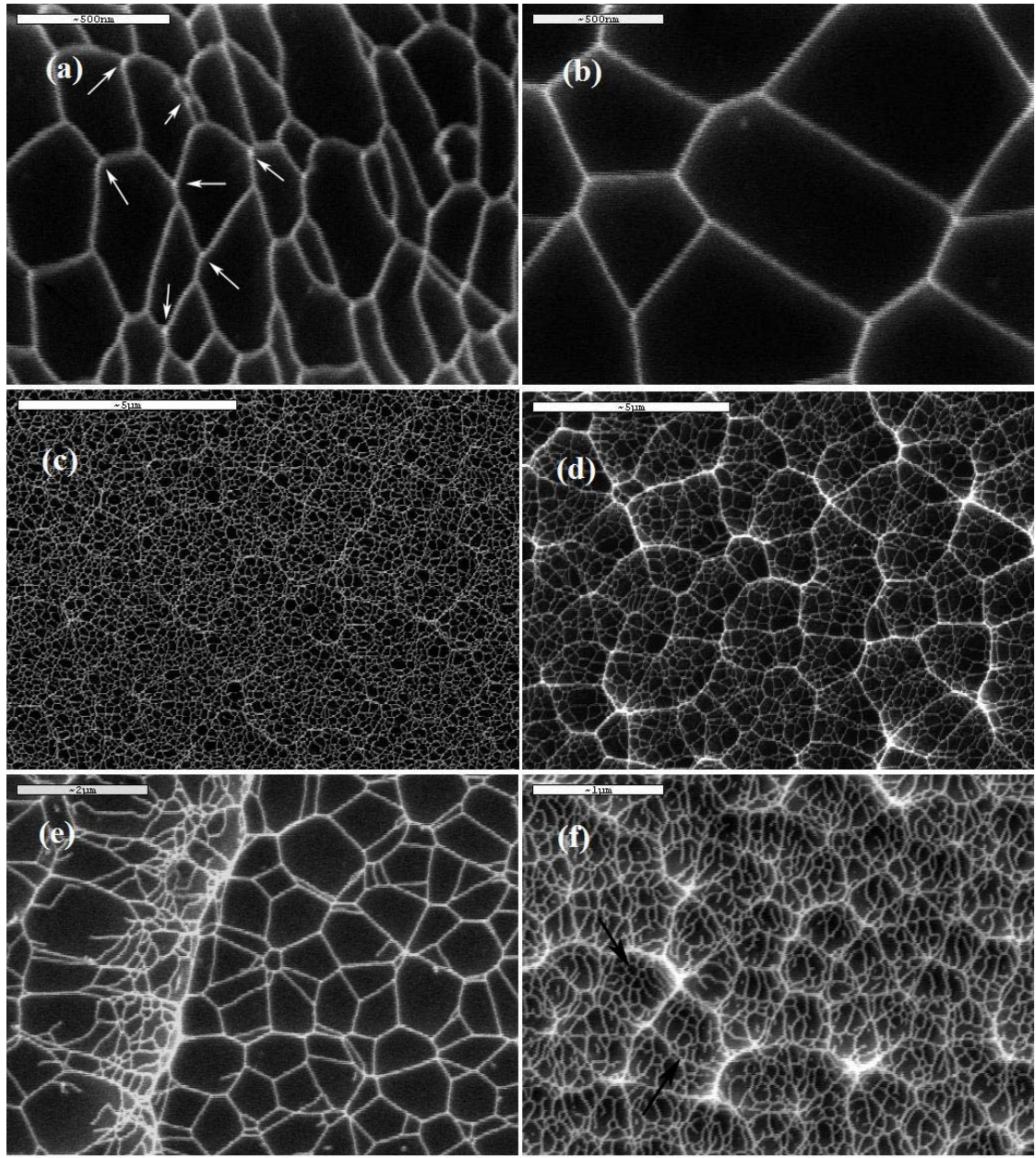

Fig. 3

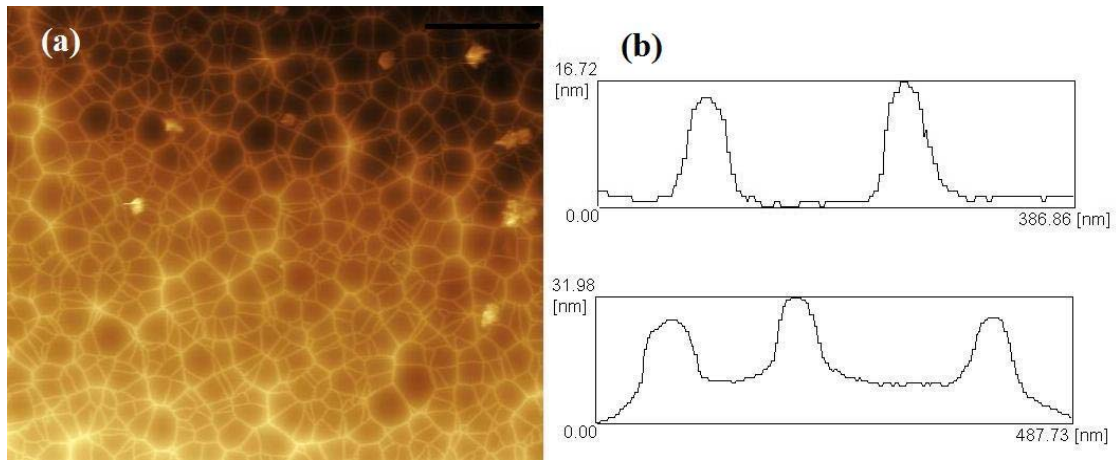

Fig. 4

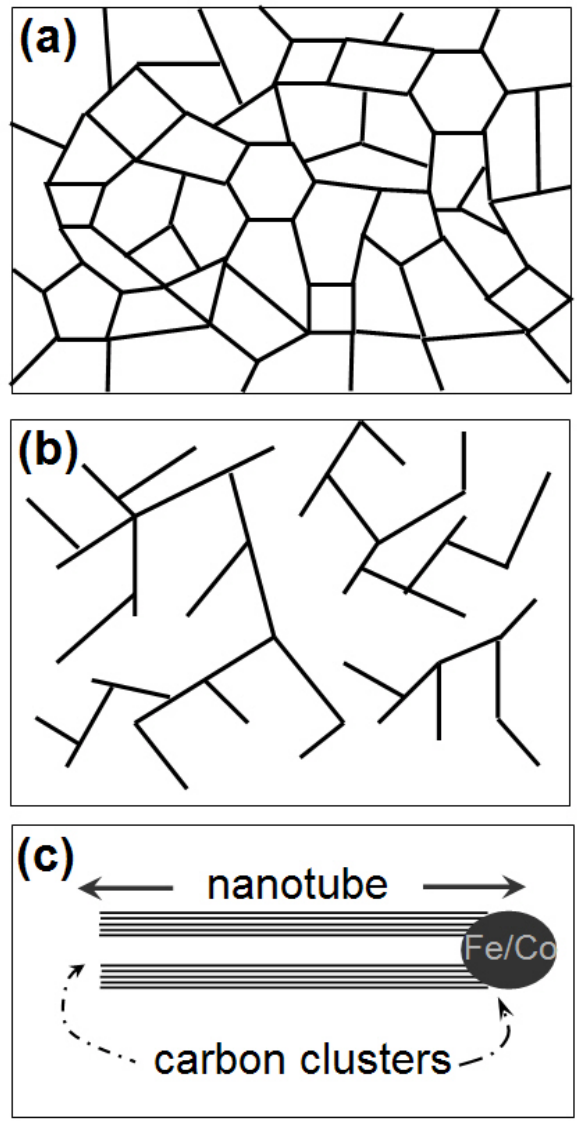